\documentclass[aps,twocolumn,showpacs,nofootinbib]{revtex4-1}

\usepackage{amsmath}
\usepackage{amsfonts,amssymb}
\usepackage{multirow}
\usepackage[dvips]{graphicx}
\usepackage{float}
\usepackage{appendix}
\usepackage{color}
\usepackage{url}
\usepackage{subfigure}
\usepackage{footnote}

\newcommand{\beq}{\begin{equation}}
\newcommand{\eeq}{\end{equation}}
\newcommand{\bea}{\begin{eqnarray}}
\newcommand{\eea}{\end{eqnarray}}

\def\nl{\nonumber\\}

\RequirePackage{xspace} \allowdisplaybreaks

\begin{document}

\def\lambdac{\Lambda_c^+}
\def\lambdacbar{\bar\Lambda_c^-}
\def\pkpipi{\Lambda_c^+\to p K^-\pi^+\pi^0}
\def\pkpipibar{\bar\Lambda_c^-\to \bar{p} K^+\pi^-\pi^0}
\def\pkspipi{\Lambda_c^+\to p K_{S}^0\pi^+\pi^-}
\def\pkspipibar{\bar\Lambda_c^-\to \bar{p} K_{S}^0\pi^-\pi^-}
\def\lamtripi{\Lambda_c^+\to \Lambda\pi^+\pi^+\pi^-}
\def\lamtripibar{\bar\Lambda_c^-\to \bar{\Lambda}\pi^+\pi^-\pi^-}
\def\CT{C_{\hat{T}}}
\def\CTbar{\bar{C}_{\hat{T}}}
\def\AT{A_{\hat{T}}}
\def\ATbar{\bar{A}_{\hat{T}}}
\def\aP{a_{\rm P}}
\def\aCP{\delta_{\rm CP}}
\def\Lam{\Lambda}
\def \cdots{\cdot\cdot\cdot}

\title{Prospects for CP and P violation in $\Lambda_{c}^+$ decays at STCF}
\author{%
Xiao-Dong Shi $^{1,2}$\email{xiaodong.shi@mail.ustc.edu.cn}%
\quad Xian-Wei Kang $^{3}$
\quad Ikaros Bigi $^4$\\
\quad Wei-Ping Wang $^{1,2}$
\quad Hai-Ping Peng $^{1,2}$
}

\affiliation{
$^1$ State Key Laboratory of Particle Detection and Electronics, Hefei 230026, China\\
$^2$ University of Science and Technology of China, Hefei 230026, China\\
$^3$ College of Nuclear Science and Technology, Beijing Normal
University, Beijing 100875, China\\
$^4$ Physics Department, University of Notre Dame du Lac, Notre
Dame, IN 46556, USA
}

\begin{abstract}

CP violation is an excellent tool for probing flavor dynamics as we learnt first with $K_L \to 2 \pi$
and later also with the weak decays of beauty mesons.
LHCb 2019 data have shown CP violation for the first time in  $D^0\to K^-K^+$ vs. $D^0\to\pi^-\pi^+$.
Searching for CP asymmetries is of great interest in charm quark sector in the Standard Model (SM) or even more beyond it.
In charm hadron decays, lots of work had focused on two-body final states, and the measurements of CP asymmetries in three- or four-body final states are rare.
Dalitz plots have shown an excellent record for three-body final states, and more results are desired for four-body ones.
In this work we study CP asymmetries in the decays $\Lambda^+_c \to p K^-\pi^+\pi^0$/$\Lambda
\pi^+\pi^+\pi^-$/$pK_S\pi^+\pi^-$, where the SM gives zero values for the first two channels, while $3.3 \times 10^{-3}$ for the last one due to $K^0 - \bar K^0$ oscillation.
We performed a fast Monte Carlo simulation study by using electron-positron annihilation data of  1~$\textrm{ab}^{-1}$ at center-of-mass energy $\sqrt{s}\, =\, 4.64$ GeV.
 The data is expected to be available by the next generation Super Tau Charm Facility proposed by China and Russia with one year (or even less) data taking operation.
The results indicate that a sensitivity at the level of $0.2\sim0.5$\% is accessible for these processes, which would be enough to measure nonzero CP-violating asymmetries as large as 1\%.
Furthermore $A_{\rm T-odd} \neq 0$ can establish parity violation (PV) by themselves and likewise for $\bar A_{\rm T-odd} \neq 0$. The SM is based on
$W^{+/-}$ being 100 \% left-handed. One can compare decay asymmetry parameters $a_{\rm P}$ from $\Lambda_c^+ \to \Lambda e^+\nu$ vs.
$\bar a_{\rm P}$ from $\bar \Lambda_c^- \to \bar \Lambda e^- \bar \nu$. In the SM one gets $a_{\rm P} =1$ \& $\bar a_{\rm P}  = -1$ in the SM,
while present data give: $(a_{\rm P} + \bar a_{\rm P} )/2 = 0.00 \pm 0.04$.
Probing these non-leptonic decays of $\Lambda_c^+$ would give new lessons about non-perturbative QCD or even indirect impact of New Dynamics on PV.

\end{abstract}

\maketitle


\section{Introduction}
\label{INTRO}

Manifestations of CP violation (CPV) predicted by the Cabibbo-Kobayashi-Maskawa
(CKM) mechanism~\cite{KM} in the Standard Model (SM) are in impressive agreement
with experimental results, especially for the strange and beauty
quark sectors~\cite{Charles:2004jd,Charles2015,Bona:2006ah}
\footnote{However, it is not big enough to account for the
matter-antimatter asymmetry which leaves one reason for searching
for New Physics (NP) beyond SM.}. CPV in the charm quark sector
predicted by the SM is at the order of $10^{-3}$ in singly
Cabibbo suppressed decays and much less for doubly Cabibbo suppressed ones~\cite{WOLF,Bigibook,AHN}. 
The level of $10^{-3}$ has been near the upper limit of the spread of a substantial range of predictions in the literature, 
and not really a typical estimate.
For the first time CPV has been shown in the weak decays of charm mesons, namely in $D^0\to K^-K^+$ vs. $D^0\to\pi^-\pi^+$ in the LHCb 2019 data~\cite{Aaij:2019kcg}.
Additionally, CPV has never been observed in the decays of baryons, except for the evidence in the $\Lambda^0_{b} \to p \pi^-\pi^+\pi^-$ decay~\cite{Aaij:2016cla}.

We point out that non-leptonic decays of charmed hadrons mostly occur with by many-body final states (FS) (and even more for beauty ones); crucial information is given there
about fundamental dynamics, not as a `background' for two-body FS.
For three-body FS decays, we have a well-known tool, namely Dalitz plots with an excellent record.
Yet one has to continue to four-body ones since we have to learn much more at least.
Furthermore, inspired by evidence for CPV in $\Lambda^0_b \to p\pi^-\pi^+\pi^-$ from LHCb data~\cite{Aaij:2016cla}, it is interesting and meaningful to study CPV by the method of triple-product asymmetries in the charmed {\em baryon} $\lambdac$ decay.

There is an obvious, but important comment. When discussing CPV in the weak decays of beauty hadrons, one mostly looks at CKM suppressed transition processes.
What about CKM favored ones? Indirect CPV has been established in the decay $B^0 \to J/\psi K_S$; the CKM favored amplitude of
$b \to c \bar c s$ gives $|V_{cb}V^*_{cs}|\sim {\cal O}( \lambda^2) \simeq 0.05 << 1$.
However, the situation is very different for charm hadrons, where the leading source is described by $|V_{cs}V^*_{ud}| \simeq 1 - \lambda^2 \simeq 0.95$. Furthermore
charm baryons can produce direct CPV only. Thus the SM can {\em not} explain sizeable CPV asymmetries for $V_{cs}V^*_{ud}$ amplitudes in general, and in particular for
$\Lambda^+_c \to p K^-\pi^+\pi^0$/$\Lambda \pi^+\pi^+\pi^-$. Yet there is a special case,
the SM predicts CPV for $\Lambda^+_c \to pK_S\pi^+\pi^-$ at `around' $3.3\, \times \, 10^{-3}$ due to CPV in $K^0 - \bar K^0$ oscillation
~\cite{Bigibook}, although it is not due to $\Delta C \neq 0$.
This similar prediction for CPV has been tested for $D^{\pm} \to K_S\pi^{\pm}$ with some success:  $A_{CP}(D^+ \to K_S \pi^+) = (-0.41 \pm 0.09)$ \%;
yet the `landscape' is more complex for $\Lambda^+_c$.
It would be close to a miracle, if new physics (NP) could produce non-zero CPV for $\Lambda^+_c \to p K^-\pi^+\pi^0$/$\Lambda \pi^+\pi^+\pi^-$ or
sizably above $3.3\, \times \, 10^{-3}$ for $\Lambda^+_c \to pK_S\pi^+\pi^-$, but it is possible.
Thus experimenters cannot ignore that. With more data and refined analyses in the future one can use much better tools to calibrate favored decays, when one goes for accuracy.
One has to be open-minded about this project. Our community has successful experience with triple-product asymmetries $A_{\rm T-odd}$ and $\bar A_{\rm T-odd}$ (see also Sec.~\ref{CPVIOL} below).
In the weak decays of charm baryons one goes after parity violation(PV) and direct CPV measurements in somewhat different ways.

$A_{\rm T-odd} \neq 0$ establish PV by itself and likewise for $\bar
A_{\rm T-odd}$: \beq \label{PV4} a_{\rm P} \equiv A_{\rm T-odd} \neq
0 \; \; \; \; \; , \; \; \; \; \;\bar a_{\rm P} \equiv \bar A_{\rm
T-odd} \neq 0\; ; \eeq in practice one can test experimental
uncertainties by comparing $ A_{\rm T-odd}$ vs. $\bar A_{\rm
T-odd}$. In the literature, e.g. in~\cite{Aaij:2016cla}, PV is also defined as $(a_{\rm P}+\bar
a_{\rm P})/2$. The SM produces large
PV; we will back to that below. As we had said above, the
`landscape' of $\Delta C \neq 0$ is close to CP invariance; thus one
can connect CV (charged conjugation violation) with PV: $a_{\rm P} +
a_{\rm C}\simeq 0$. Using different words to describe the same
situations we know that these CP asymmetries are very small at best:
\beq \label{CPV4} \delta_{\rm CP} \equiv \frac{1}{2}(A_{\rm T-odd} -
\bar A_{\rm T-odd}) \ll 1 \; . \eeq Strong final state interactions
(FSI) are not the source of CPV. That has to come from new dynamics
(ND) with weak phases -- yet FSI should show their impact. One has
to be realistic: very likely we will not find CPV in these weak
decays of $\Lambda_c^+$. Yet it is {\em not} a waste of time, and those channels 
are worth to do in experiment due to the following points: 

\begin{itemize}
 \item It is not a miracle to find CPV in Cabibbo suppressed decays of $\Lambda_c^+$; one can use those mentioned channels to calibrate Singly Cabibbo suppressed (SCS) decays to probe {\em regional} CP asymmetries in $\Lambda_c^+ \to p\pi^-\pi^+\pi^0$/$pK^-K^+\pi^0$/$\Lambda K^+\pi^+\pi^-$ with accuracy in the future.
\item One expects sizable PV in the weak decays of $\Lambda_c^+$.
\item At least, one can get novel lessons about the impact of strong forces close to thresholds, namely about non-perturbative QCD.
\end{itemize}
We will consider three decay processes : $\lamtripi$, $\pkpipi$ and $\Lambda^+_c \to
pK_S\pi^+\pi^-$ due to their large branching fractions~\cite{PDG}: \footnote{Maybe one can
measure also BR $(\Lambda^+_c \to pK_L\pi^+\pi^-) \sim 1.6 \% $.}
\bea {\rm BR}(\pkpipi) &\simeq& \; 4.4 \; \% \; \;, \nl {\rm
BR}(\lamtripi) &\simeq& \; 3.6\; \% \; \; ,\nl {\rm BR}(\Lambda^+_c
\to pK_S\pi^+\pi^-) &\simeq& \; 1.6 \; \% . \eea  The current paper
is mainly dedicated to the study of physics sensitivities that can be achieved at a future Super Tau Charm Facility (STCF), where the central values for PV and CPV quantities of charmed baryon decays are surely measurable.

The new generation STCF is an electron-positron collider to operate at the $\tau$-charm energy region, with
peak luminosity above $0.5\times 10^{35} \; \textrm{cm}^{-2}\textrm{s}^{-1}$ at a center-of-mass
energy (CME) of $\sqrt{s}\sim$4~GeV/$c^2$~\cite{Bonder,IPAC2018,PengCharm2018}. The facility is discussed strongly
and proposed by the Chinese and Russian high energy physics communities in last few years, and is expected to
be realized in the coming ten years. With such high luminosity, the proposed STCF can deliver electron-position collisions to accumulate more than 1~ab$^{-1}$ of integrated luminosity per year, thus providing an excellent opportunity to study charm physics, notably including CPV with charmed meson and baryon decays.

In the electron-positron annihilation process, the $\Lambda_c$ baryon can be produced via the process $e^+e^-\to\Lambda_c^+\bar{\Lambda}_c^-$ abundantly. The Belle experiment has measured the production cross section of $e^+e^-\to\Lambda_c^+\bar{\Lambda}_c^-$ by the initial state radiation (ISR) process, where a peak is depicted as the measured Born cross section with the value of $\sigma$$\sim$~470~pb at CME of $\sqrt{s}$=4.63~GeV/$c^2$, and is assigned to be originated from the charmonium-like state $Y(4630)$ decay~\cite{Pakhlova:2008vn}.
The BESIII experiment has collected data at CME of $\sqrt{s}=$4.6~GeV/$c^2$ with integrated luminosity of 567~pb$^{-1}$, as well as other three data sets at lower CME but above the $\Lambda_c^+\bar{\Lambda}_c^-$ mass threshold ($\sqrt{s}$=4.575, 4.580 and 4.590 GeV) with more than one order smaller luminosity (47.7, 8.54, and 8.16~pb$^{-1}$, respectively).
With these data sets, BESIII is very productive, and has published several interesting results, such as the production cross section of $e^+e^-\to \Lambda_c^+\bar{\Lambda}_c^-$, the absolute decay branching fractions of $\Lambda_c^+\to pK^-\pi^+$ as well as other eleven Cabibbo favored (CF) hadronic modes, the branching fractions of SCS decays, the decay with neutron included, semi-leptonic decay and inclusive decays $etc$~\cite{Ablikim:2017ors, Ablikim:2016mcr, Ablikim:2016tze, Ablikim:2015prg, Ablikim:2017lct, Ablikim:2015flg}.
In proton-proton collisions, such as at the LHCb experiment, the $\Lambda_c^+$ baryon is abundantly produced directly from proton-proton collision or via beauty baryon decays~\cite{Aaij:2017nsd,Aaij:2017xva,Aaij:2017rin}.
Comparing to the Belle II and LHCb experiments, the STCF is of shortage in statistics. However, STCF has several advantages, such as the excellent ratio of signal to background, the perfect detection efficiency,
the well controlled systematic uncertainty and the capability of full event reconstruction, $etc$.
By implementing the double tag (DT) method, STCF can perform systematic researches of $\Lambda_c^+$ decays, including the absolute measurements of semi-leptonic decays and the decays with a neutron, $K_L$ or invisible particles included in final state~\cite{Li:2016tlt}.
Besides studying $\Lambda_c^+$ physics, STCF will play crucial role in the study of
how the $Y(4630)$ state enters $e^+e^-\to \Lambda_c^+\bar\Lambda_c^-$ production~\cite{DaiLY}, the mixing of axial-vector
mesons~\cite{KangSemiD}, the proton form factors~\cite{Kangppbar1,Kangppbar2}, $etc$.

In what follows, we will perform a careful investigation for the sensitivities on CPV and PV in the decays $\lamtripi$, $\pkpipi$ and $\Lambda^+_c \to pK_S\pi^+\pi^-$ at
the future STCF. There is a rich `landscape' about
strong and weak forces; one needs more refined analyses -- but we
have the tools for that; all we need is more data.

\section{Observables}

The situations between PV \& CPV are very different as said above; thus the goals are also different. The first example:
with more data one should find non-zero values of PV in these non-leptonic transitions.

\subsection{Parity asymmetries}
\label{PARITYVIOL}

It had been realized that it is a crucial test of the SM: charged $W^{\pm}$ bosons are left-handed, as we had learnt from
$\pi^+/K^+ \to \mu^+ \nu$ vs. $\pi^+/K^+ \to e^+ \nu$; so far, we have not seen
right-handed one. 2018 PDG data \cite{PDG} have shown PV in $\Lambda_c^+ \to \Lambda l^+\nu$ that are consistent with the SM predictions,
although with sizable uncertainties:
\beq
(a_{\rm P} + \bar a_{\rm P})/2 = 0.00 \pm 0.04  \; .
\eeq
On the other hand, this situation is not well tested in
non-leptonic decays. Probing these non-leptonic decays of $\Lambda_c^+$ would give new lessons about non-perturbative QCD or even indirect impact of New Dynamics on
PV. In these non-leptonic decays of $\Lambda_c^+$ T-odd moments should produce sizable PV with different values, see the
Eq.(\ref{PV4}). We have added these analyses of PV below.
Indeed, one gets a non-trivial test of this experiment.

One should expect large values of PV in those non-leptonic transitions. A small/tiny value of PV would be signal of NP. However,
one cannot predict future results of PV even within the SM.  It means our community would learn new lessons about the impact of strong forces.
So far, no true predictions can be given due to non-perturbative QCD with many resonances in the region of  0.5 - 2 GeV, including broad ones like
$f_0(500)$, $K^*_0(700)$ etc. Our main paint is that we describe the travel to use, when our community has the future data to get the information about
the underlying dynamics.

\subsection{CP asymmetries}
\label{CPVIOL}

The SM predicts tiny CPV in charm baryon decays; therefore large statistics
are required. Obviously one goes after direct CPV. The landscape of
data is very `flat' for CP asymmetries: it is expected to be very unlikely
that any evidence for CPV in Cabibbo favored(CF) transitions is found, e.g.
$\Lambda_c^+ \to \Lambda \pi^+$~\cite{FOCUS} or
$\Lambda_c^+ \to \Lambda e^+ \nu$~\cite{CLEO}
, where CPV was investigated by measuring the decay asymmetry
parameters. There are also recent theoretical papers about the decay
asymmetry parameters~\cite{Cheng2018Lambdac,LvXiaorui,PingRG}, and
especially in Ref.~\cite{Cheng2018Lambdac} the model calculations
are done for the singly Cabibbo-suppressed decays. We also notice
that BESIII has measured the absolute branching fraction of
$\Lambda_c^+ \to \Lambda l^+ \nu$ with less
uncertainties~\cite{Ablikim:2015prg}. We exploit triple-product
asymmetries composed by four-momenta with{\em out} recurring to the
information of polarization as has been done in
Refs.~\cite{KangDDbar,Kang:2009iy}.

The CF decays of $\Lambda_c^+$ baryons with multi-hadrons in final state, such as  $\pkpipi$, $\lamtripi$ and $\Lambda^+_c \to pK_S\pi^+\pi^-$ depict a much `complex' landscape~\cite{PDG,Ablikim:2015flg},
which is believed to give us much more
information about the underlying dynamics than that of
$\Lambda_c^+ \to \Lambda \pi^+$ and $\Lambda_c^+ \to \Lambda e^+\nu$, but need both
more data \& refined analyses. To describe the four-body weak decays
of $\Lambda_c^+$ one has one baryon in the FS, $p$ or $\Lambda$,
plus three pseudoscalar mesons -- kaons or pions. In the rest frame of the
charm baryon we have two observables of spin-1/2 -- $
\textbf{s}_{\rm \Lambda_c}$ and $\textbf{s}_{p/\Lambda}$ -- and the
momenta of the four particles -- $\textbf{p}_{p/\Lambda}$ plus the momenta
of the three mesons. One can describe T-odd moments in different
ways, which give us the same information about the
underlying dynamics; however with finite data and lack of perfect
control of QCD, some are better than others.
We exploit the scalar triple products to construct CPV observables, see the
Refs.~\cite{Kang1,Kang:2009iy,Kang:2010td,Bevan:2014nva,Gronau:2015gha,Valencia,Datta1,
Datta2,Rosner1,Rosner2,Durieux:2016nqr,Durieux:2015zwa}.
These papers came mostly from theorists who had
focused on singly Cabibbo transitions. This method has been widely applied in several experiments, see
recent ones in Refs.~\cite{Aaij:2016cla,Prasanth:2017beu,Aaij:2018lsx,Aaij:2016nki}.
Some early ones can be found in Refs.~\cite{delAmoSanchez:2010xj,Aaij:2014qwa,Lees:2011dx}.

For these $\Lambda_c^+$ decays, the scalar triple products
$\CT =\textbf{p}_{p/\Lambda}\cdot(\textbf{p}_{h_1}\times\textbf{p}_{h_2})$ and the conjugate, $\CTbar = \textbf{p}_{\bar p/\bar \Lambda}\cdot(\textbf{p}_{\bar h_1}\times\textbf{p}_{\bar h_2})$,
with pseudo-scalar mesons $h_i$, are defined to study CPV. The momenta
$\textbf{p}$ are measured in the rest frame of the $\Lambda^+_c$ baryon; When two $\pi^+$ (or two $\pi^-$) mesons are present, that one with the larger momentum is selected.
The asymmetries are then defined as:
\begin{eqnarray}
\label{eq:AT}
\AT(\CT)&=&\frac{N(\CT>0)-N(\CT<0)}{N(\CT>0)+N(\CT<0)}\;\;, \nl
\ATbar(\CTbar)&=&\frac{\bar{N}(-\CTbar>0)-\bar{N}(-\CTbar<0)}{\bar{N}(-\CTbar>0)+\bar{N}(-\CTbar<0)}.
\end{eqnarray}
These correspond to $A_{\rm T-odd}$ $\bar A_{\rm T-odd}$ moments;
CPV observables are $\delta_{\rm CP}$, see Eq.~(\ref{CPV4}). Any
significant deviation from zero indicates CPV; in particular, one
also looks for the number $N$ of events for the {\it direct} CPV asymmetries:
\begin{eqnarray}\label{eq:DCP}
A_{CP}^{(K)}&=&\frac{N(\pkpipi)-N(\pkpipibar)}{N(\pkpipi)+N(\pkpipibar)}\;,\nl
A_{CP}^{(\pi)}&=&\frac{N(\lamtripi)-N(\lamtripibar)}{N(\lamtripi)+N(\lamtripibar)}\;,\nl
A_{CP}^{(K_{S}^{0})}&=&\frac{N(\pkspipi)-N(\pkspipibar)}{N(\pkspipi)+N(\pkspipibar)}\;,\nl
\end{eqnarray}
One can expect sizable values of $\AT$ and $\ATbar$ due to
FSI effects~\cite{Bigibook,Valencia}. It is
also possible to find non-zero CPV. In Ref.~\cite{Kang:2010td}, the authors show that large CPV can indeed happen
in NP with the two-Higgs doublet model as an example. The CP violation $\sim 0.18\sin\phi$ with $\phi$ denoting 
the New-Physics CP violating phase. Then it can reach 18\% if $\sin\phi$ is close to 1. 

The measurements may vary over the phase
space due to resonant contributions or their interference effects,
which may be cancelled if integrating over the whole phase space.
For the decays $\pkpipi$, $\lamtripi$ and $\pkspipi$,
the semi-regional CPV is measured with respect to several bins separated by the
dihedral angle, and Monte Carlo (MC) simulation is also exploited to study this case.
We stress again that no CP asymmetry has been found yet
in the transitions of charm baryons. Therefore, one has to probe CPV
with more data and tools, although this is not trivial.

\section{Measurement Procedure}
\label{Sec:ExpPro}

The STCF project is in research and development stage. To maximize the physics potential, a BESIII-like detector but with much improved performance for each sub system is proposed.
From inside to outside, the STCF detector consists of a tracking system, a particle identification (PID) system, a high granularity electromagnetic calorimeter (EMC) and a muon detector with high $\mu$/$\pi$ separation capability.
To be competitive on high precision measurements, and to cope with high
event rate and radiation dose, several advanced technologies are proposed to be
the STCF sub-detectors, such as a thin silicon detector or a micro pattern gas detector for the inner tracking system,
a Cherenkov based PID system, crystal LYSO or pure CsI based electromagnetic calorimeter $etc$.
To investigate the physics potential capability and optimize the detector design, a fast simulation tool dedicated to the STCF detector has been developed, where the detection efficiency and measurement resolution of each sub-detector are parameterized according to an empirical formula and the BESIII detector performance, and the parameters are adjustable flexibly. The event generators for both signal and background processes are migrated from the BESIII experiment.
The tool has been validated by the BESIII full simulation package~\cite{Li:2009ay} using {\sc Geant4}, and provides a perfect platform to perform physics studies with huge statistics. A note dedicated to this tool is under preparation.

To study the sensitivities of CPV and PV in the decays $\lamtripi$, $\pkpipi$ and $\Lambda^+_c \to pK_S\pi^+\pi^-$, both signals and inclusive MC samples are generated based on the STCF fast simulation tool, where the parameters for each sub-detector are from BESIII.
In this study, the $\Lambda^+_c$ signal is originated from the process $e^+e^-\to\Lambda_c^+\Lambda_c^-$ at the CME of $\sqrt{s}$=4.64~GeV, where the peak of the production cross section lies.
The study is performed based on the integrated luminosity of 1~ab$^{-1}$, which is expected to be achieved at STCF within one year (or even less) of data taking.
In the simulation, $e^+e^-$ collisions are simulated by the KKMC generator~\cite{kkmc}, which takes into account the beam energy spread and the ISR correction, where the beam energy spread is assigned to be same value as that of BEPCII.
To study the potential background and optimize event selection, an inclusive MC sample, which includes $\lambdac\lambdacbar$ pair production, $l^+l^-$ ($l = e,\mu,\tau)$ events, open charm processes, ISR-produced low-mass $\psi$ states, and the continuum process $e^+e^-\rightarrow q\bar{q}$ with $q=u, d, s$ quarks~\cite{Ping:2008zz} are generated with the integrated luminosity of 1~ab$^{-1}$, where the decays of intermediate states, such as $\Lambda_c^+$ baryons, charmed mesons, charmonium state, and light hadrons, is performed according to the branching fractions quoted from PDG.
To study the signal shape and detection efficiency, the signal MC samples of $\pkpipi$, $\lamtripi$ and $\Lambda_c^+ \to p
K_S\pi^+\pi^-$ are generated with uniform distribution in phase space; no intermediate state in the two or three bodies is considered.
The real data will show the impact of intermediate states, such as $\rho$, $K^*$, $\Delta$ $etc$.

\begin{figure}[!h]
\begin{center}
\subfigure[]{
\includegraphics[width=8cm]{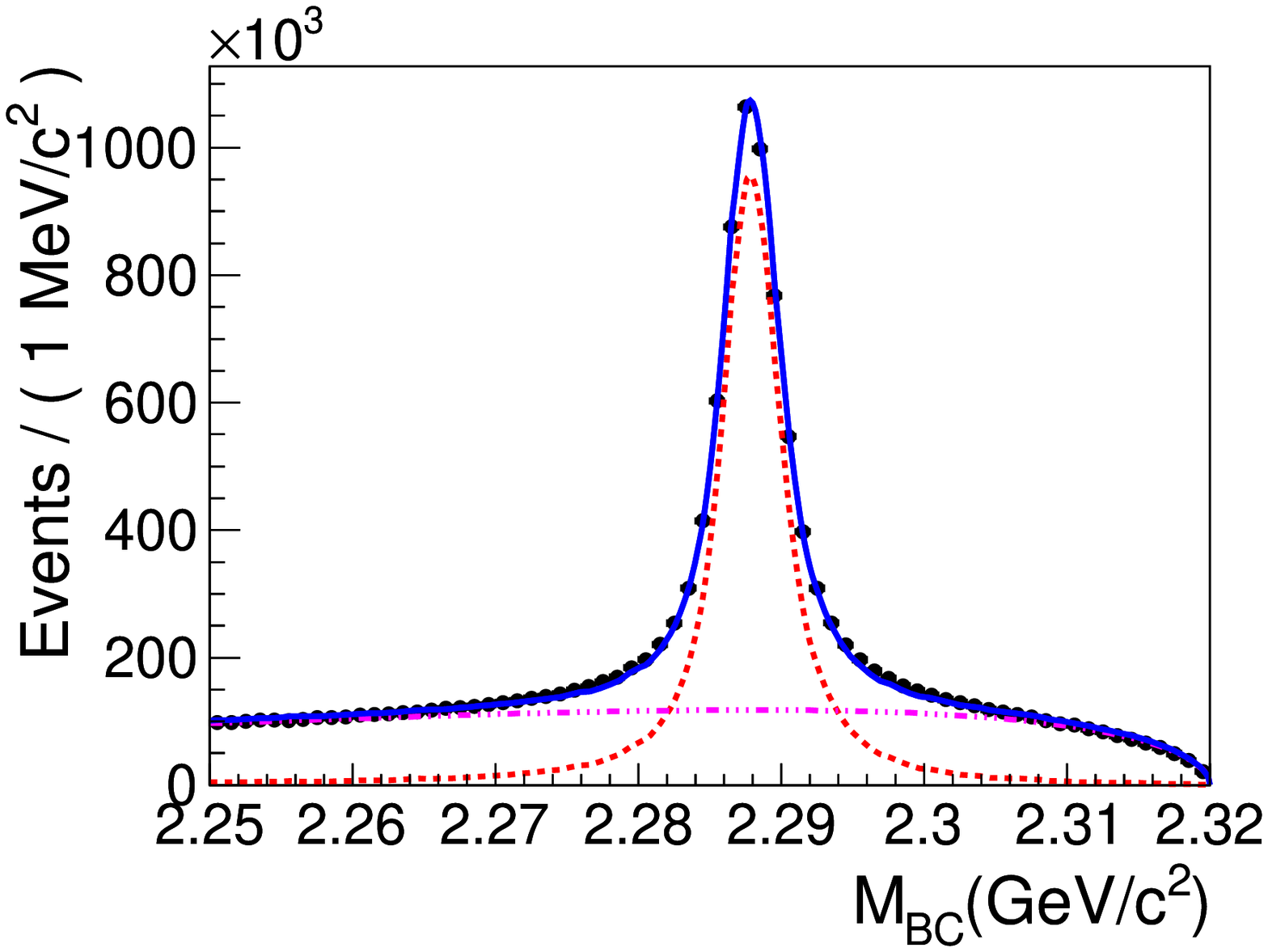}}
\subfigure[]{
\includegraphics[width=8cm]{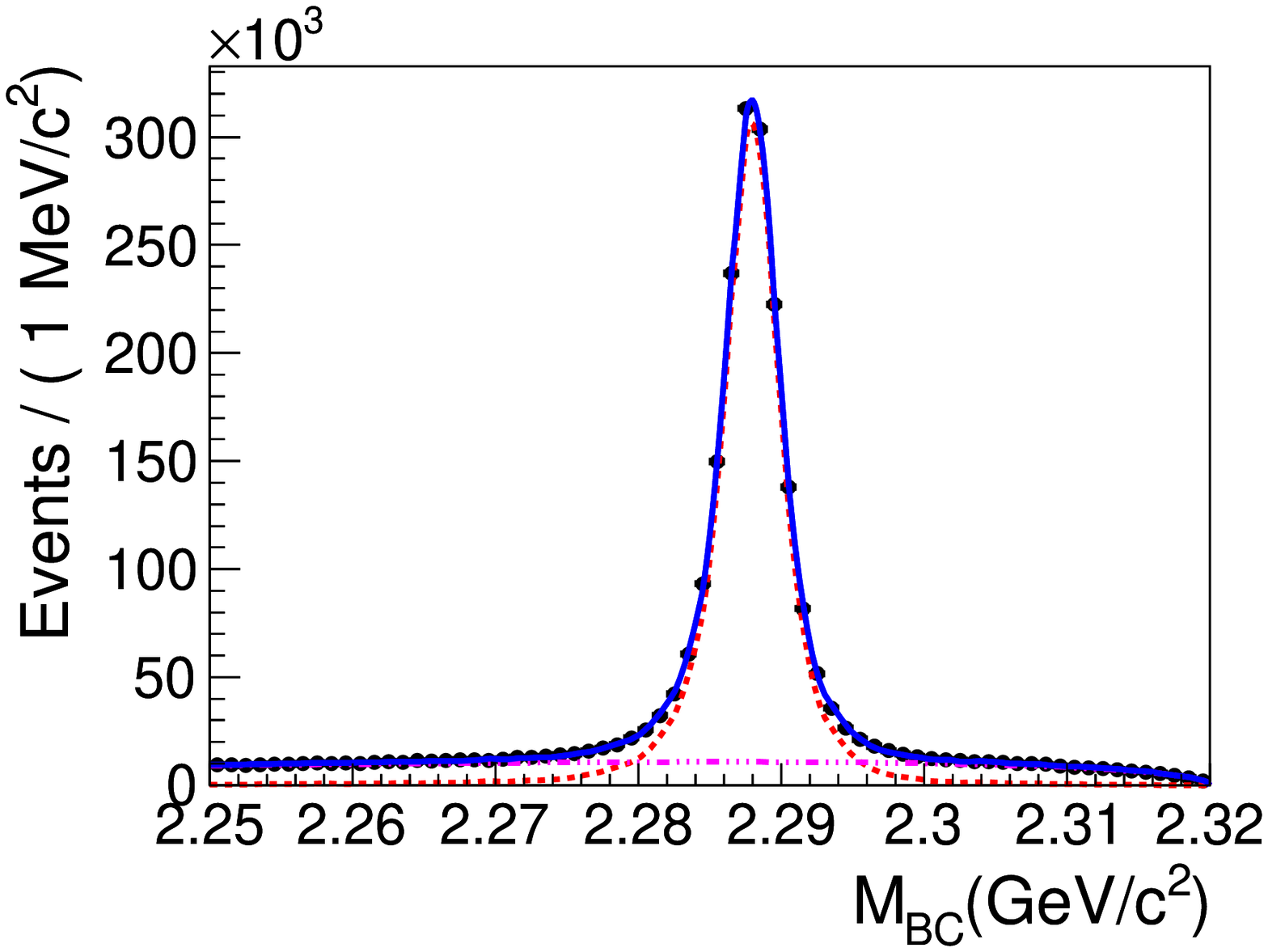}}
\subfigure[]{
\includegraphics[width=8cm]{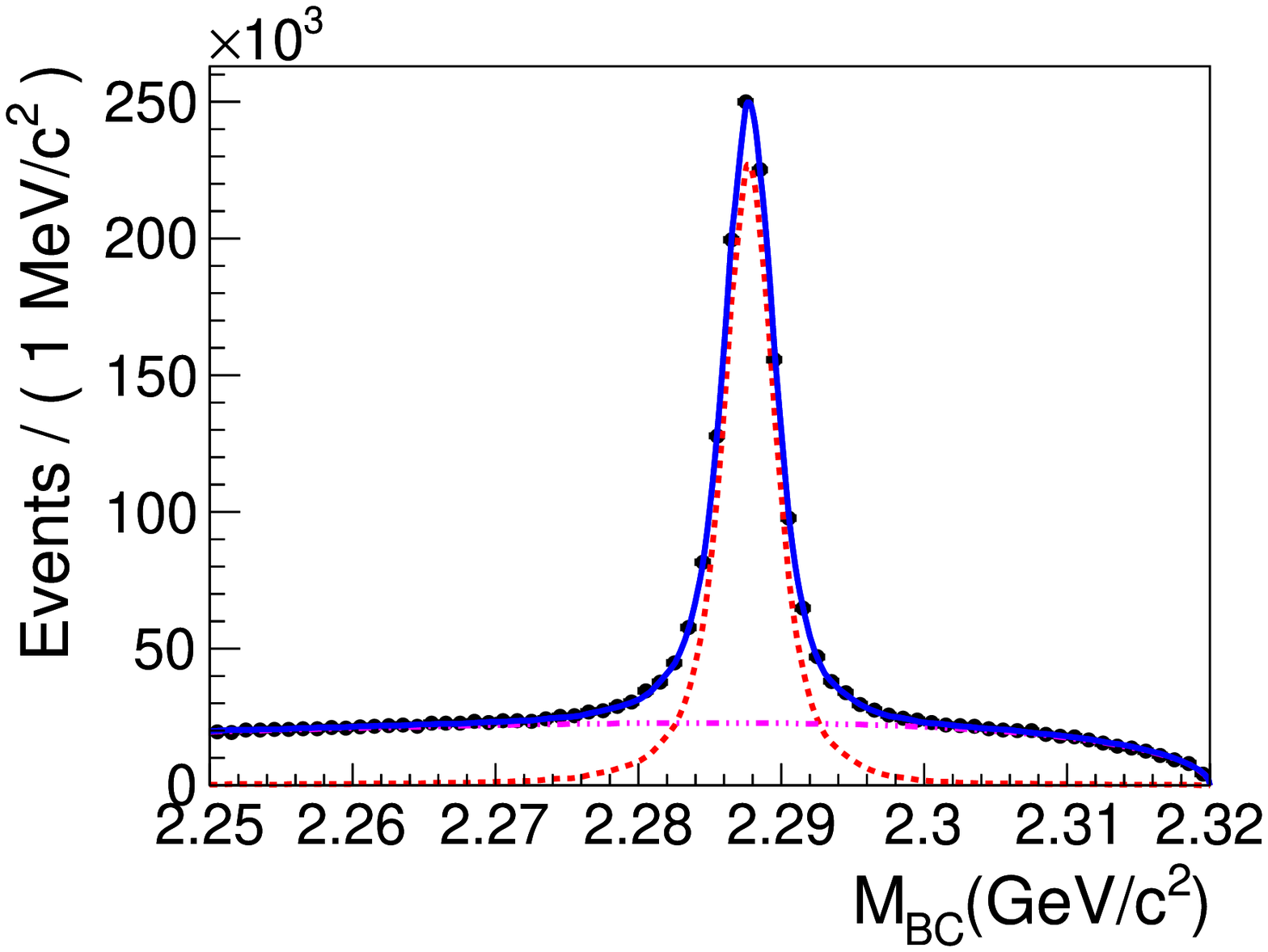}}
\caption{\label{mbc}$M_{BC}$ distribution for (a) $\pkpipi$, (b)
$\lamtripi$ and (c) $\pkspipi$ decays. The dots correspond to the MC simulation.
The blue solid curves are the fit funtions, while the pink
dotted (red dashed) lines represent the backgrounds (signals).}
\end{center}
\end{figure}

In this analysis, the single tag method is implemented to improve the statistics.
Candidate events are selected with the similar criteria (including charged tracks, $\pi^0$ and $K_S$ candidates selection, PID, $etc$.) as in Ref.~\cite{Ablikim:2017lct} according to the final state of signal.
The signal yields are dertermined by performing a binned maximum likelihood fit to the distribution of the beam constrained mass $M_{BC}$, which is defined as $M_{BC}\equiv\sqrt{E^2_{\rm{beam}}/c^4-{p_{\Lambda_c}^2/c^2}}$, with $E_{\rm {beam}}$ denoting the energy of the electron/positron beam and $p_{\Lambda_c}$ the three-momentum of the $\lambdac$ candidate calculated from the momenta of the final-state particles in the initial $e^+e^-$ center-of-mass system.
Figure~\ref{mbc} shows the $M_{BC}$ distributions for $\pkpipi$, $\lamtripi$ and $\Lambda_c^+ \to p K_s\pi^+\pi^-$ decays corresponding to 1~ab$^{-1}$ of an inclusive MC sample, where $\Delta E$, defined as $\Delta E=E_{{\rm beam}} - E_{\Lambda_c}$ with $E_{\Lambda_c}$ denoting the energy of $\Lambda_c$ candidate summing over the energy of the corresponding final state particles, is required to be within three times of its resolution.
Clear $\Lambda_c^+$ signals with low background are observed.
Detailed studies by the inclusive MC sample indicate that there is no peaking background in the $M_{BC}$ distributions. Thus, in the fit to determine the signal yields, the shape of background is described by an ARGUS function~\cite{Albrecht:1990am} with fixed high-end truncation, and those of signal are obtained from the signal MC samples.

For semi-regional CPV, one may discretize the dihedral angle and/or the invariant mass into different bins, as in Ref.~\cite{Aaij:2016cla}.
In the intermediate state regions, strong phases are enhanced and thus can provide opportunity for large CP asymmetries due to large interference.
Since the components of intermediate states are unknown due to the lack of experimental data, in this study, we split the phase space into different bins along the dihedral angle $\Phi$ distribution only, and the binning along the invariant mass distribution is not considered.
Here, $\Phi$ is the angle between the decay planes formed by the $p\pi^0$ and $K^-\pi^+$ ($p\pi^-$ and $K_S^0\pi^+$, $\Lambda\pi^+_{\rm{fast}}$ and $\pi^+_{\text{slow}}\pi^-$)
for the process $\pkpipi$ ($\pkspipi$,\,$\lamtripi$).
In the future, once collecting huge data at STCF, we can have a better understanding of the underlying dynamics of the $\Lambda^+_c$ decay, including the impact of broad intermediate states, such as $K^*_0(700)/\kappa$ and $f_0(500)/\sigma$ $etc$ , and the analyses of semi-regional CPV can be refined\footnote{ We also note that the knowledge of
the two-photon couplings to the scalars~\cite{Dai2gamma} is helpful
to understand their structures.}.

\section{Results and discussions}
Following the approaches described in Sec.~\ref{Sec:ExpPro}, we report in Table \ref{tab:direct} the physics sensitivities for direct CPV, as defined in Eq.~\eqref{eq:DCP}, as well as for PV and CPV observables constructed from the $T$-odd moments elaborated in Eqs.~\eqref{PV4} and~\eqref{CPV4}. The physics sensitivities include the statistical uncertainties only; systematic uncertainties are expected to be well under control \footnote{Only the systematic uncertainty related with the asymmetry between positive and negative charged tracking will have to be taken into account.}.
By error propagation, according to Eqs.~\eqref{eq:DCP},~\eqref{PV4} and~\eqref{CPV4}, if we ignore the impact of the statistical uncertainty from background contamination, and assume $N_{\Lambda_c^+}=N_{\bar\Lambda_c^-}=N$ and $N(\CT>0)=N(\CT<0)=\bar{N}(\CTbar>0)=\bar{N}(\CTbar<0)=N/2$, the statistical uncertainties for $A_{CP}$, ($\aP$+$\bar a_{\rm{P}}$)/2 and $\aCP$ are $1/\sqrt{2N}$, where $N_{\Lambda_c^+}$ and $N_{\bar\Lambda_c^-}$ are the numbers of $\Lambda_c^+$ and $\bar\Lambda_c^-$ candidate events, and $N(\CT>0)$ ($\bar{N}(\CTbar>0)$) and $N(\CT<0)$ ($\bar{N}(\CTbar<0)$) are the numbers of candidate events with $\CT>0$ and $\CT<0$ for the $\Lambda_c^+$ ($\bar\Lambda_c^-$) candidates, respectively.
Thus, as shown in Table~\ref{tab:direct}, the three measured variables have the same sensitivities, mostly due to the small impact from the background, and provide complementary and more comprehensive information to search for PV and CPV in $\Lambda_c^+$ hadronic decays.
With an $e^+e^-\to\Lambda_c^+ \Lambda_c^-$ data sample of 1~ab$^{-1}$ integrated luminosity at $\sqrt{s}$ = 4.64~GeV collected by STCF, the physics sensitivities to search for PV and CPV are at the few permille level for three interesting decay modes, individually, which are at the level of potential CPV in charm sector and unambiguous PV if observed.
\begin{table}[!h]
\centering
\caption{The physics sensitivities for direct CPV as well as $(\aP+\bar a_{\text{P}})/2$ (PV) and $\aCP$ (CPV) constructed from the T-odd moments for $\pkpipi$, $\lamtripi$ and $\pkspipi$ processes with 1~$\textrm{ab}^{-1}$ of data at $\sqrt{s}$ = 4.64~GeV at STCF.
The results for $(\aP+\bar a_{\text{P}})/2$ and $\aCP$ are the same.
} \label{tab:direct}
\begin{tabular*}{0.9\linewidth}{@{\extracolsep{\fill}}ccc}
\hline \hline
channel &   direct CPV & $(\aP+\bar a_{\text{P}})/2$,\,\,$\aCP$\\
\hline\hline
$\pkpipi$   &0.0025  &0.0026\\
$\lamtripi$ &0.0052  &0.0052\\
$\pkspipi$  &0.0040  &0.0041\\
\hline\hline
\end{tabular*}

\end{table}

As discussed previously, the sensitivity on CPV may be enlarged in some regions of phase space due to the enhancement of the strong phase and interference.
This kind of CPV is called semi-regional CPV or localized CPV, and is of great interest for both theorists and experimentalists.
In this study, we also perform a sensitivity study for semi-regional CPV for the three $\Lambda_c^+$ decay models, individually.
Due to the lack of information on the intermediate states, the studies are performed only by binning the dihedral angle $\Phi$, as defined in Sec.~\ref{Sec:ExpPro}, based on MC samples generated with a phase-space model.
The measurements with real data are expected to be of better sensitivity due to the contribution from intermediate states.
In this study, we discretize the dihedral angle $\Phi$ into ten bins with equal steps from 0 to $\pi$, and measure the T-odd moments CPV in each bin.
As shown in Table~\ref{tab:local}, the sensitivities for  $\Lambda_c^+\to p K^-\pi^+\pi^0$, $\Lambda\pi^+\pi^+\pi^-$ and $p K_{S}^0\pi^+\pi^-$ in each bin are around 0.0080,\, 0.016,\, and 0.013, respectively, which are smaller by a factor 1/$\sqrt{10}$ relative to global CPV values, since the statistics is reduced by a factor 10 in each bin.

\begin{table}[!h]
\centering
\caption{The sensitivities for semi-regional CPV ($\aCP$) in $\pkpipi$, $\lamtripi$ and $\pkspipi$  decays with ten bins, for 1~ab$^{-1}$ at $\sqrt{s}$=4.64~GeV at STCF.} \label{tab:local}
\begin{tabular*}{0.9\linewidth}{@{\extracolsep{\fill}}cccc}
\hline \hline
$\Phi$ & $p K^-\pi^+\pi^0$ & $\Lambda\pi^+\pi^+\pi^-$ & $p K_{S}^0\pi^+\pi^-$ \\
\hline\hline
(0, 0.1$\pi$)&	0.0078   & 0.016 & 0.013\\
(0.1$\pi$, 0.2$\pi$)& 	0.0080   & 0.016 & 0.013\\
(0.2$\pi$, 0.3$\pi$)&  0.0081   & 0.017 & 0.013\\
(0.3$\pi$, 0.4$\pi$)&  0.0082   & 0.017 & 0.013\\
(0.4$\pi$, 0.5$\pi$)&  0.0083   & 0.017 & 0.013\\
(0.5$\pi$, 0.6$\pi$)&  0.0083   & 0.017 & 0.013\\
(0.6$\pi$, 0.7$\pi$)&  0.0083   & 0.017 & 0.013\\
(0.7$\pi$, 0.8$\pi$)&  0.0080   & 0.016 & 0.013\\
(0.8$\pi$, 0.9$\pi$)&  0.0079   & 0.016 & 0.013\\
(0.9$\pi$, $\pi$)&  0.0077   & 0.016 & 0.013\\
\hline\hline
\end{tabular*}

\end{table}

\section{Conclusions and prospects}
Searching for CPV and PV in charmed baryon decays certainly provide complementary and comprehensive information to understand the underlaying dynamics of charmed hadrons and test the SM, and is of great interest both for theorists and experimentalists.
The future Super Tau Charm Facility (STCF) proposed by Chinese and Russian scientists may provide great platform for these kinds of studies due to its characters of high luminosity, broad center-of-mass energy acceptance, abundant production, clean environment, $etc$.
In this work, we propose to study direct CPV by measuring the asymmetries of decay branching fractions between charge conjugate modes as well as PV and CPV by constructing $T-$odd moments in $\Lambda_c^+$ decays to multi-hadron final states.
We study the physics sensitivities for CPV and PV in the decays $\pkpipi$, $\lamtripi$ and $\Lambda^+\to p K_S\pi^+\pi^-$ by performing a fast simulation, where the $\Lambda^+_c$ is assumed to be from the $e^+e^-$ annihilation to  $\Lambda_c^+\bar{\Lambda}_c^-$ pair
at center-of-mass energy of $\sqrt{s}$ = 4.64~GeV with  1~ab$^{-1}$ $e^+e^-$ integrated luminosity, \emph{i.e.} expected to be available in one year (or even less) operating at future STCF.
The results indicate that the physics sensitivities are around 0.25$\sim$0.5\% for the three decay modes, individually, which is at the level of potential CPV in charm hadron sector or for an unambiguous PV observation.
We also discuss how semi-regional CPV may be enlarged due to the enhancement of the strong phase and interference,
and perform the sensitivity study for the same decay modes by binning the dihedral angle distribution.
Simulations cannot give predictions, in particular for many-body final states.
In the future, with huge real data collected at STCF, we can also study the intermediate states and their impact.
Many exciting results are expected at STCF, providing excellent information for non-perturbative QCD studies.

\section*{Acknowledgment}
The authors Xiao-Dong Shi, Wei-Ping Wang and Hai-Ping Peng thank Hefei Comprehensive National Science Center and the Supercomputing Center of USTC for their strong support.
The authors also gratefully appreciate for the useful discussions with Prof.Hai-Bo Li.
This work is supported by the Double First-Class University Project Foundation of USTC and
the National Natural Science Foundation of
China under Projects No. 11275199, No.11805012, No.11625523;
the National Science Foundation under the grant number PHY-1820860.

\end{document}